**Quantifying nonadiabaticity in major families of superconductors**

Evgeny F. Talantsev[1,2*]

[1]M.N. Miheev Institute of Metal Physics, Ural Branch, Russian Academy of Sciences, 18, S. Kovalevskoy St., Ekaterinburg, 620108, Russia

[2]NANOTECH Centre, Ural Federal University, 19 Mira St., Ekaterinburg, 620002, Russia

## Abstract

The classical Bardeen-Cooper-Schrieffer and Eliashberg theories of the electron–phonon-mediated superconductivity are based on the Migdal theorem, which is an assumption that the energy of charge carriers, $k_B T_F$, significantly exceeds the phononic energy, $\hbar\omega_D$, of the crystalline lattice. This assumption, which is also known as adiabatic approximation, implies that the superconductor exhibits fast charge carriers and slow phonons. This picture is valid for pure metals and metallic alloys because these superconductors exhibit $\frac{\hbar\omega_D}{k_B T_F} < 0.01$. However, *n*-type doped semiconducting $SrTiO_3$ was the first superconductor which beyond this adiabatic approximation, because this material exhibit $\frac{\hbar\omega_D}{k_B T_F} \cong 50$. There is growing number of newly discovered superconductors which also beyond the adiabatic approximation. Here, leaving apart pure theoretical aspects of nonadiabatic superconductors, we classified major classes of superconductors (including, elements, A-15 and Heusler alloys, Laves phases, intermetallics, noncentrosymmetric compounds, cuprates, pnictides, highly-compressed hydrides and oxygen, and magic-angle twisted bilayer graphene) by the strength of nonadiabaticity (for which the ratio of the Debye temperature to the Fermi temperature, $\frac{T_\theta}{T_F}$, is used as a criterion for the nonadiabaticity) versus the superconducting transition temperature, $T_c$. The discussion of this classification scheme and its relation to other known classification counterparts is given.

**Quantifying nonadiabaticity in major families of superconductors**

## 1. Introduction

The majority of experimental works in superconductivity utilizes classical the Bardeen-Cooper-Schrieffer (BCS) [1] and the Migdal-Eliashberg (ME) [2,3] theories as primary tool to analyse measured data. However, as this was pointed out by Pietronero and coworkers [4], that these theories are valid for superconductors which satisfy the condition designated as Born-Oppenheimer-Migdal approximation [4]:

$$\frac{\hbar\omega_D}{k_B T_F} = \frac{T_\theta}{T_F} = \left.\frac{88\ K}{1.1\times10^5\ K}\right|_{Pb} = 8\times10^{-4} \ll 1, \qquad (1)$$

where $\hbar$ is reduced Planck constant, $\omega_D$ is the Debye frequency, $k_B$ the Boltzmann constant, $T_\theta$ is the Debye temperature, $T_F$ is the Fermi temperature, and data for lead reported by Poole [5]. Born-Oppenheimer-Migdal approximation allows to separate electronic and ionic motions in metals, because Eq. 1 implies that the conductor exhibits fast charge carriers (for which characteristic energy scale is related to the Fermi temperature, $T_F$) and relatively slow phonons (for which characteristic energy scale is related to the Debye temperature, $T_\theta$).

However, Eq. 1 satisfies for many, but not for all superconductors, and the first discovered superconductor for which Eq. 1 was found to be violated is *n*-type doped semiconducting $SrTiO_3$ [6]:

$$\frac{\hbar\omega_D}{k_B T_F} = \frac{T_\theta}{T_F} = \left.\frac{627\ K}{13\ K}\right|_{SrTiO_3} = 48 \gg 1, \qquad (2)$$

where data for $SrTiO_3$ is taken from [7,8]. Theoretical description of the superconductivity in materials, in which the charge carriers and the lattice vibrations exhibit similar to Eq. 2 characteristic energy scales is complicated and general designation of this superconductors is nonadiabatic superconductors [9-16]. The theory [9-16] provides general equation for the superconducting transition temperature, $T_c$, in nonadiabatic superconductors [9]: $T_c = 1.134 \times \frac{\varepsilon_F}{k_B} \times e^{-\frac{1}{\lambda_{nad}}}$, where $\varepsilon_F$ is the Fermi energy, and $\lambda_{nad}$ is the coupling strength constant in nonadiabatic superconductors, which serves similar role to the electron-phonon coupling strength, $\lambda_{e-ph}$, in the BCS [1] and ME [2,3] theories. And one of the primary fundamental theoretical problem is to calculate this constant with acceptable accuracy to describe the experiment [4-16].

For experimentalists, it is important to have simple practical routine to establish the strength of nonantiadiabatic effects in newly discovered superconductor. The most obvious parameter, which can be served as experimentally measured value to quantify the strength of nonantiadiabaticity is the $\frac{T_\theta}{T_F}$ ratio. For practical use of this criterion, there is a need for the taxonomy of possible $\frac{T_\theta}{T_F}$ values.

In this paper we analysed experimental data for major families of superconductors and based on our analysis we proposed the following classification scheme:

$$\begin{cases} \frac{T_\theta}{T_F} < 0.025 \rightarrow adiabatic\ superconductor; \\ 0.025 \lesssim \frac{T_\theta}{T_F} \lesssim 0.4 \rightarrow moderate\ nonadiabaticity; \\ 0.4 < \frac{T_\theta}{T_F} \rightarrow\ nonadiabatic\ superconductor; \end{cases} \quad (3)$$

One of our findings is that for weakly nonadiabatic superconductors (i.e. for materials exhibited $0.025 \leq \frac{T_\theta}{T_F} \lesssim 0.4$) the predicting power of the BCS-ME theories (for instance, the prediction of the superconducting transition temperature) is reasonably accurate. However, all these superconductors are located outside of the BCS corner in the Uemura plot.

We also showed how the proposed classification scheme is linked to some other known empirical scaling laws and taxonomies in superconductivity [13,17-21], while the search for the link of the proposed taxonomy with recently reported big data [22,23] is under a progress.

## 2. Utilized models

Proposed taxonomy is based on the knowledge of three fundamental temperatures of the superconductor, which are $T_c$, $T_\theta$, and $T_F$. The superconducting transition temperature, $T_c$, is directly measured in either temperature resistance, either in magnetization, experiments, and it is important to mention primary experimental techniques and theoretical models utilized to deduce the Debye temperature, $T_\theta$, and the Fermi temperature, $T_F$, in superconductors.

There are two primary techniques to determine the Debye temperature, $T_\theta$. One technique is to analyse measured the temperature dependent normal-state specific heat, $C_p(T)$, from which the electronic specific heat coefficient, $\gamma_n$, and the Debye temperature, $T_\theta$, are deduced (see, for instance [24-26]):

$$\frac{C_p(T)}{T} = \gamma_n + \beta T^2 + \alpha T^4 \qquad (4)$$

where $\beta$ is the Debye law lattice heat-capacity contribution, and $\alpha$ is from higher-order lattice contributions. The Debye temperature can be calculated using:

$$T_\theta = \left(\frac{12\pi^4 Rp}{5\beta}\right)^{\frac{1}{3}} \qquad (5)$$

where $R$ is the molar gas constant, and $p$ is the number of atoms per formula unit.

Another technique is to fit normal-state temperature dependent resistance, $R(T)$, to the Bloch-Grüneisen (BG) equation [24-28]:

$$R(T) = \frac{1}{\frac{1}{R_{sat}} + \frac{1}{R_0 + A \times \left(\frac{T}{T_\theta}\right)^5 \times \int_0^{\frac{T_\theta}{T}} \frac{x^5}{(e^x-1)(1-e^{-x})} \cdot dx}} \qquad (6)$$

where, $R_{sat}$ is the saturated resistance at high temperatures which is temperature independent, $R_0$ is the residual resistance at $T \to 0\ K$, and $A$ is free fitting parameter. Many research groups utilized both techniques (i.e. Eqs. 4-6) to deduce $T_\theta$ [24-27,29].

From measured $T_c$ and deduced $T_\theta$, one can derived the electron-phonon coupling constant, $\lambda_{e-ph}$, as a root of either original McMillan equation [30], either its recently revisited form [27]:

$$T_c = \left(\frac{1}{1.45}\right) \times T_\theta \times e^{-\left(\frac{1.04(1+\lambda_{e-ph})}{\lambda_{e-ph}-\mu^*(1+0.62\lambda_{e-ph})}\right)} \times f_1 \times f_2^* \quad (7)$$

$$f_1 = \left(1 + \left(\frac{\lambda_{e-ph}}{2.46(1+3.8\mu^*)}\right)^{3/2}\right)^{1/3} \quad (8)$$

$$f_2^* = 1 + (0.0241 - 0.0735 \times \mu^*) \times \lambda_{e-ph}^2 \quad (9)$$

where $\mu^*$ is the Coulomb pseudopotential, $0.10 \lesssim \mu^* \lesssim 0.15$ [27,30].

There are several experimental techniques to derive the Fermi temperature, $T_F$, from experimental data. One of these techniques is to measure temperature dependent Seebeck coefficient, $S(T)$, and fit measured dataset to the equation [8]:

$$\left|\frac{S(T)}{T}\right| = \frac{\pi^2}{3} \frac{k_B}{e} \frac{1}{T_F} \quad (10)$$

Another approach is to measure the magnetic quantum oscillations [31] from which the magnitude of charge carrier mass, $m^* = m_e(1 + \lambda_{e-ph})$ (where $m_e$ is bare mass od electron), together with the size of the Fermi wave vector, $k_F$, can be obtained and plugged into [31]:

$$T_F = \frac{\hbar^2}{2k_B} \frac{k_F^2}{m^*} \quad (11)$$

Alternative approach is based on the extracting of the charge carriers mass, $m^*$, and density, $n$, as two of four parameters from the simultaneous analysis of $C_p(T)$, $R(T)$, the muon spin relaxation ($\mu$SR), the lower critical field data, $B_{c1}(T)$, and the upper critical field data, $B_{c2}(T)$ [32], and plugging these parameters into equation for isotropic spherical Fermi surface [32]:

$$T_F = \frac{\hbar^2}{2k_B} \frac{1}{m^*} (3\pi^2 n_s)^{\frac{2}{3}} \quad (12)$$

where $n_s$ is bulk charge curriers density at $T \to 0\ K$. For 3D superconductors, $n_s$ is given by the equation [33]:

$$n_s(0) = \frac{m^*}{\mu_0 e^2}\frac{1}{\lambda^2(0)}, \quad (13)$$

where $\mu_0$ is the permeability of free space, $l$ is the charge carrier mean free path, $\lambda(0)$ is the ground state London penetration depth, and $\xi(0)$ is the ground state coherence length.

It should be noted, that $\lambda(0)$ can be also deduced from the ground state lower critical field [28,34]:

$$B_{c1}(0) = \frac{\phi_0}{4\pi}\frac{ln\left(1+\sqrt{2}\kappa(0)\right)}{\lambda^2(0)}, \quad (14)$$

where $\kappa(0) = \frac{\lambda(0)}{\xi(0)}$ is the ground state Ginzburg-Landau parameter.

For two dimensional (2D) superconductors, $T_F$ can be determined from $\mu$SR measurements and crystallographic data [18]:

$$T_F = \frac{\pi\hbar^2}{k_B}\frac{1}{m^*}n_s \times c_{int}, \quad (15)$$

where $c_{int}$ is the average distance between superconducting planes.

If measuring techniques are limited to the magnetoresistance measurements, $R(T,B)$ (which was the case in the field of highly-compressed near-room temperature superconductors (NRTS) [35-50], until recent experimental progress by Minkov *et al* [51,52]), $T_F$ can be estimated by the equation [53]:

$$T_F = \frac{\pi^2 m^*}{2k_B\hbar^2} \times \xi^2(0) \times \Delta^2(0), \quad (16)$$

where $\Delta(0)$ is the ground state amplitude of the superconducting energy gap, which is varying in reasonably narrow range $3.2 \leq \frac{2\Delta(0)}{k_B T_c} \leq 5.0$, that the ballpark value for $T_F$ can be estimated.

## 3. Results

In Table 1 we presented data for major groups of superconductors, where data sources for $T_c$, $T_\theta$, $T_F$, and other parameters (for instance, $\lambda_{e-ph}$) are given.

In Figure 1we showed $T_c$ vs $T_F$ dataset in log-log plot, which is traditional data representation in well-known Uemura plot [18].

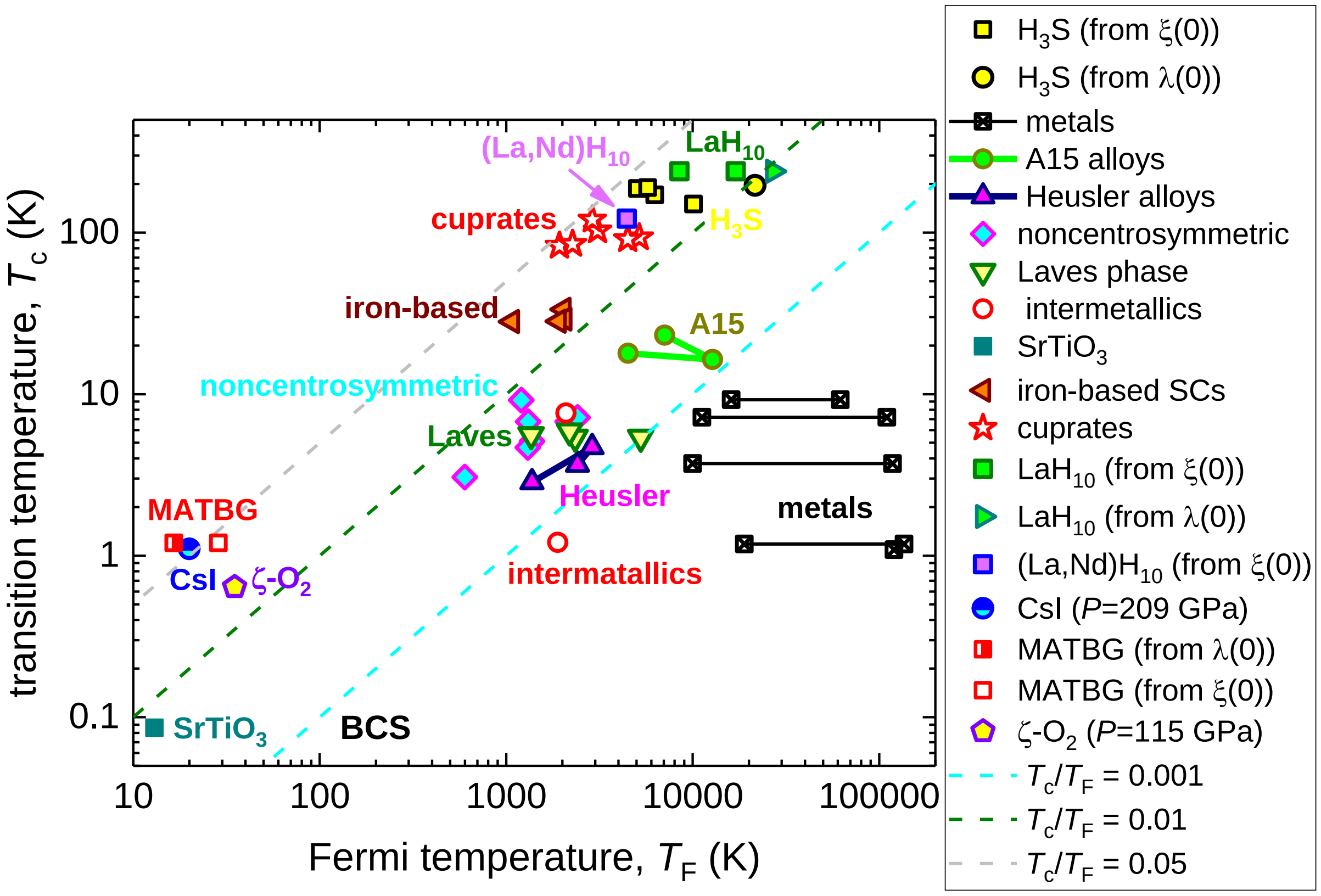

**Figure 1.** Uemura plot ($T_c$ vs $T_F$) for primary superconducting families. References on original data ($T_c$ and $T_F$) can be found in Table 1.

In Figure 2 we represented same superconducting materials, but here we displayed $\lambda_{e-ph}$ vs $\frac{T_\theta}{T_F}$ dataset in semi-log plot. For our best knowledge, $\lambda_{e-ph}$ vs $\frac{T_\theta}{T_F}$ plot was first plotted by Pietronero *et al* [13] in linear-linear scales. However, because $\frac{T_\theta}{T_F}$ ratio for main families of superconductors is varied within four orders of magnitude (Table 1), and $0.4 \leq \lambda_{e-ph} \leq 3.0$ it is more suitable to use semi-log plot (Fig. 2).

**Table 1.** Superconductors and their parameters used in the work. In all calculations (except some original sources), $\mu^* = 0.13$.

| **Type / Chemical composition** | **λ(0) (nm)** | **ξ(0) (nm)** | **$\lambda_{e\text{-}ph}$** | **$T_c$ (K)** | **$T_\theta$ (K)** | $\frac{2\Delta(0)}{k_B T_c}$ | **$T_F$ ($10^3$ K)** | **$T_\theta/T_F$** |
|---|---|---|---|---|---|---|---|---|
| ***Pure metals*** | | | | | | | | |
| Aluminium | | | | 1.18 [58] | 394 [5] | | 136 [5] | $2.9 \times 10^{-3}$ |
| Aluminium | 50 [57] | 1550 [58] | 0.43 [59] | 1.18 [58] | 394 [5] | 3.535 [59] | 18.9 (Eq. 12) | $2.1 \times 10^{-2}$ |
| Tin | | | | 3.72 [58] | 170 [5] | | 118 [5] | $1.4 \times 10^{-3}$ |
| Tin | 77 [60] | 180 [58] | 0.72 [59] | 3.72 [58] | 170 [5] | 3.705 [59] | 10.0 (Eq. 12) | $1.2 \times 10^{-2}$ |
| Lead | | | | 7.20 [58] | 88 [5] | | 110 [5] | $8 \times 10^{-4}$ |
| Lead | 64 [60] | 87 [58] | 1.55 [59] | 7.20 [60] | 88 [5] | 4.497 [59] | 11.2 (Eq. 12) | $7.8 \times 10^{-3}$ |
| Niobium | | | | 9.25 [58] | 265 [5] | | 61.8 [5] | $4.3 \times 10^{-3}$ |
| Niobium | 52 [58] | 39 [58] | 0.98 [59] | 9.25 [58] | 265 [5] | 3.964 [59] | 16.1 (Eq. 12) | $1.6 \times 10^{-2}$ |
| Gallium | | | 2.25 [59] | 1.09 [120] | 325 [121] | | 120 [120] | $2.7 \times 10^{-3}$ |
| ***A15 Alloys*** | | | | | | | | |
| $Nb_3Sn$ | 124 [61] | 3.6 [61] | 1.8 [62] | 17.9 [61] | 234 [61] | 4.2 [62] | 4.5 (Eq. 12) | $5.2 \times 10^{-2}$ |
| $V_3Si$ | 62 [62] | 3.3 [62] | 0.96 [62] | 16.4 [62] | 297 [64] | 3.7 [62] | 12.8 (Eq. 12) | $2.3 \times 10^{-2}$ |
| $Nb_3Ge$ | 90 [58] | 3.0 [58] | 1.60 [59] | 23.2 [58] | 302 [65] | 4.364 [59] | 7.1 (Eq. 12) | $4.3 \times 10^{-2}$ |
| ***Heusler alloys*** | | | | | | | | |
| $ZrNi_2Ga$ | 350 [66] | 15 [66] | 0.551 [66] | 2.85 [66] | 300 [66] | | 1.4 (Eq. 12) | $2.2 \times 10^{-1}$ |
| $YPd_2Sn$ | 196 [67] | 19 [67] | 0.70 [67] | 4.7 [67] | 210 [67] | 4.1 [67] | 2.9 (Eq. 12) | $7.2 \times 10^{-2}$ |
| $HfPd_2Al$ | 225 [67] | 13 [67] | 0.68 [67] | 3.66 [67] | 182 [67] | 3.74 [67] | 2.4 (Eq. 12) | $7.5 \times 10^{-2}$ |
| ***Noncentrosymmetric*** | | | | | | | | |
| $Nb_{0.5}Os_{0.5}$ | 654 [68] | 7.8 [68] | 0.53 [68] | 3.07 [68] | 367 [68] | 3.62 [68] | 0.60 (Eq. 12) | $6.1 \times 10^{-1}$ |
| $Re_6Zr$ (mSR) | 356 [29] | 3.7 [29] | 0.67 [29] | 6.75 [29] | 338 [29] | 3.72 [29] | 1.3 | $2.6 \times 10^{-1}$ |
| $Re_6Zr$ (magnetization) | 247 [29] | 3.3 [29] | 0.67 [29] | 6.75 [29] | 237 [29] | 3.72 [29] | 2.1 | $1.1 \times 10^{-1}$ |
| $Mo_3Al_2C$ | 376 [69] | 4.2 [69] | 0.74 (Eqs. 7-9) | 9.2 [69] | 339 [69] | 4.03 [69] | 1.2 | $2.8 \times 10^{-1}$ |
| $NbIr_2B_2$ [70] | 223 | 4.5 | 0.74 | 7.18 | 274 | | 2.4 | $1.1 \times 10^{-1}$ |
| $TaIr_2B_2$ [70] | 342 | 4.7 | 0.70 | 5.1 | 230 | | 1.4 | $1.7 \times 10^{-1}$ |
| $Re_3Ta$ [71] | | | 0.62 | 4.7 | 321 | | 0.64 | $5.0 \times 10^{-1}$ |
| ***Laves phases*** | | | | | | | | |
| $BaRh_2$ [72] | 340 | 8.4 | 0.80 | 5.6 | 178 | | 1.4 | $1.3 \times 10^{-1}$ |
| $SrRh_2$ [72] | 229 | 9.1 | 0.71 | 5.4 | 237 | | 2.3 | $1.0 \times 10^{-1}$ |
| $SrRh_2$ [73] | 121 | 8.6 | 0.93 | 5.4 | 250 | | 5.3 | $4.7 \times 10^{-2}$ |
| $SrIr_2$ [74] | 237 | 7.5 | 0.84 | 5.9 | 180 | | 2.3 | $8.2 \times 10^{-2}$ |
| ***Intermetallics*** | | | | | | | | |
| $MgCNi_3$ [75] | 248 | 4.6 | 0.74 (Eqs. 7-9) | 7.6 | 284 | | 2.1 | $1.4 \times 10^{-1}$ |
| $RuAl_6$ [76] | 265 | 27.7 | 0.81 | 1.21 | 458 | | 1.9 | $2.4 \times 10^{-1}$ |
| ***Perovskite*** | | | | | | | | |
| $SrTiO_3$ | | | 0.2 [7] | 0.086 [8] | 690 [77] | | $1.3 \times 10^{-2}$ [8] | $5.3 \times 10^{1}$ |
| ***Pnictides*** | | | | | | | | |
| ThFeAsN | 375 [78] | | 1.48 [78] | 28.1 [78] | 332 [79] | | 0.47 (Eq. 15) $c_{int} =$ 8.5 Å [78] | $7.0 \times 10^{-1}$ |
| $KCa_2Fe_4As_4F_2$ | 230 [80] | | 1.59 [80] | 33.4 [80] | 366 [80] | | 1.3 (Eq. 15) $c_{int} =$ 8.5 Å [80] | $2.9 \times 10^{-1}$ |
| $RbCa_2Fe_4As_4F_2$ | 232 [80] | | 1.45 [80] | 29.2 [80] | 332 [80] | | 1.2 (Eq. 15) $c_{int} =$ 8.5 Å [80] | $2.8 \times 10^{-1}$ |
| $CsCa_2Fe_4As_4F_2$ | 244 [80] | | 1.44 [80] | 28.3 [80] | 344 [80] | | 1.1 (Eq. 15) $c_{int} =$ 8.5 Å [80] | $3.1 \times 10^{-1}$ |
| ***Cuprates*** | | | | | | | | |
| $YBa_2Cu_3O_7$ [81] | 115 [81,82] | 2.5 [81] | 1.5 [83] | 93.2 [81] | 437 [7] | | 3.4 (Eq. 15) $c_{int} =$ 5.8 Å [83] | $1.2 \times 10^{-1}$ |
| $(Y,Dy)Ba_2Cu_3O_7$ [88] | 128 [88,89] | 2.5 [81] | 1.5 [83] | 90.4 [88,89] | 437 [7] | 4.24 [88,89] | 2.9 (Eq. 15) $c_{int} =$ 5.8 Å [83] | $1.5 \times 10^{-1}$ |
| $Bi_2Sr_2CaCu_2O_8$ [90] | 196 [89] | 1.2 [89] | | 82.7 [89] | 240 [7] | 3.9 [89] | 1.2 (Eq. 15) $c_{int} = 6$ Å [83] | $2.0 \times 10^{-1}$ |
| $Tl_2Ba_2CaCu_2O_8$ [91] | 179 [89] | 1.2 [89] | | 103 [89] | 425 [87] | 4.3 [89] | 1.5 (Eq. 15) $c_{int} = 6$ Å [83] | $2.9 \times 10^{-1}$ |
| $HgBa_2CaCu_2O_8$ [92] | 188 [89] | 1.6 [89] | | 120 [89] | 525 [87] | 3.3 [89] | 1.3 (Eq. 15) $c_{int} = 6$ Å [83] | $3.9 \times 10^{-1}$ |
| $Bi_2Sr_2Ca_2Cu_3O_{10}$ [93] | 175 [89] | 1.0 [89] | | 85 [89] | 319 [7] | 4.5 [89] | 1.5 (Eq. 15) $c_{int} = 6$ Å [83] | 2.1 $\times 10^{-1}$ |
| ***Twisted graphene*** | | | | | | | | |
| MATBG [94] | 2180 [95] | | $\frac{m^*}{m_e} = 0.2$ [95] | 1.2 [95] | 1864 [96] | 4.4 [95] | $16.5 \times 10^{-3}$ (Eq. 15) $c_{int} = 1\ nm$ | $1.1 \times 10^2$ |
| MATBG [94] | | 61.4 [95] | $\frac{m^*}{m_e} = 0.2$ [95] | 1.2 [95] | 1864 [96] | 4.4 [95] | $28.6 \times 10^{-3}$ (Eq. 16) | $6.5 \times 10^1$ |
| ***Ionic Salt*** | | | | | | | | |
| CsI (*P* = 206 GPa) [97] | | | 0.445 [98] | 1.1 [97] | 339 [98] | | $(20 \pm 4) \times 10^{-2}$ [98] | $17 \pm 4$ [98] |
| ***NRTS hydrides*** | | | | | | | | |
| $H_3S$ (*P* = 155 GPa) [54] | 37 [52] | 1.9 [54] | 2.2 [99] | 197 [99] | 1427 [99] | | 21.6 (Eq. 12) | 6.6 $\times 10^{-2}$ |
| $H_3S$ (*P* = 155 GPa) [54] | | 1.86 [100] | 1.76 [55,56,100] | 190 [100] | 1427 [99] | 3.5 [100] | 6 [100] (Eq. 16) | $(2.4) \times 10^{-1}$ |
| $H_3S$ (*P* = 155 GPa) [54] | | 1.92 [100] | 1.76 [55,56,100] | 188 [100] | 1427 [99] | 3.3 [100] | 5 [100] (Eq. 16) | $2.9 \times 10^{-1}$ |
| $H_3S$ (*P* = 160 GPa) [54] | | 2.18 [100] | 1.76 [55,56,100] | 171 [100] | 1427 [99] | 3.4 [100] | 10.7 [100] (Eq. 16) | $1.3 \times 10^{-1}$ |
| $H_3S$ (*P* = 160 GPa) [54] | | 2.67 [100] | 1.76 [55,56,100] | 150 [100] | 1427 [99] | 4.0 [100] | 10.1 [100] (Eq. 16) | $1.4 \times 10^{-1}$ |
| $LaH_{10}$ (*P* = 150 GPa) [36] | 30 [51] | | 2.77 [27] | 240 [27] | 1310 [27] | | 27.0 (Eq. 12) | 2.7 $\times 10^{-2}$ |
| $LaH_{10}$ (*P* = 150 GPa) [36] | | 1.5 [51] | 2.77 [27] | 240 [27] | 1310 [27] | 3.53 [assumed] | 8.5 (Eq. 16) | 1.5 $\times 10^{-1}$ |
| $LaH_{10}$ (*P* = 150 GPa) [36] | | 1.5 [51] | 2.77 [27] | 240 [27] | 1310 [27] | 5.0 [assumed] | 17.0 (Eq. 12) | 7.7 $\times 10^{-2}$ |
| $La_{1-x}Nd_xH_{10}$ (x=0.15) (*P* = 180 GPa) [48] | | 2.3 [101] | 1.65 [101] | 122 [101] | 1156 [101] | 4.0 [101] | 4.4 [101] (Eq. 16) | 2.6 $\times 10^{-1}$ |
| ***Compressed oxygen*** | | | | | | | | |
| ζ-$O_2$ (*P*=115 GPa) [102] | | 42 [103] | 0.42 [103] | 0.64 [103] | 306 [103] | | $3.5 \times 10^{-2}$ [103] (Eq. 16) | 8.7 |

And finally in Figure 3 we represented same superconducting materials, but here we displayed $T_c$ vs $\frac{T_\theta}{T_F}$ dataset in log-log plot. This type of plot was chosen because as $T_c$, as $\frac{T_\theta}{T_F}$ are varied within several orders of magnitude.

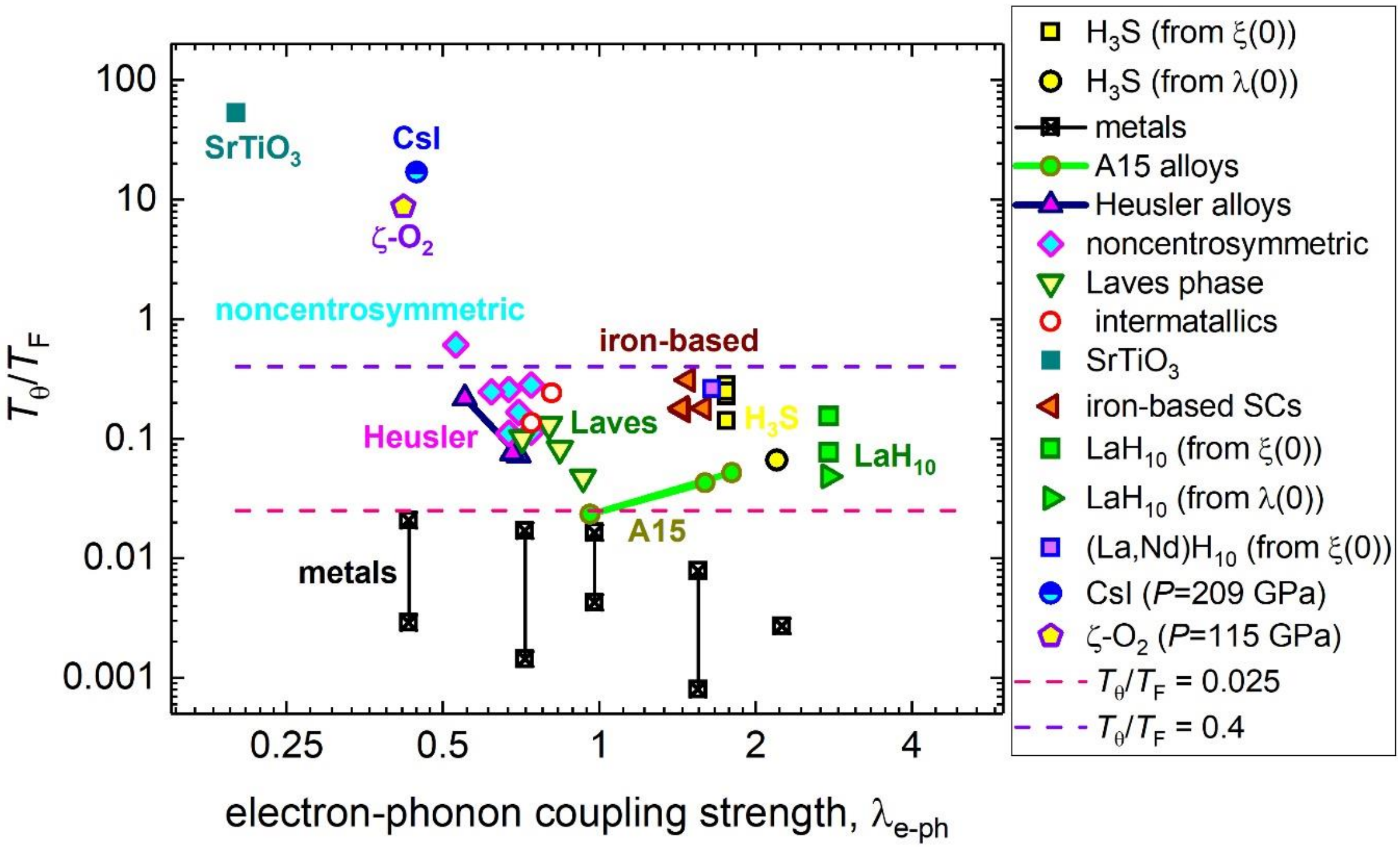

**Figure 2.** Plot of $\frac{T_\theta}{T_F}$ vs $\lambda_{e-ph}$ for primary superconducting families. This type of plot proposed by Pietronero *et al* [13]. References on original data ($T_\theta$, $\lambda_{e-ph}$, $T_F$) can be found in Table 1.

## 4. Discussion

The family of near-room temperature superconductors (NRTS) is represented in Table 1 and Fig. 1 by $H_3S$ ($P$ = 155,160 GPa), $LaH_{10}$ ($P$ = 150 GPa) and $La_{1-x}Nd_xH_{10}$ (x = 0.09, $P$ = 180 GPa). Two different approaches to calculate $T_F$ in NRTS materials were used:

1. $T_F$ was calculated based on Eqs. 12,13. In these calculations $\lambda(0) = 37\ nm$ (extracted from the analysis of DC magnetization experiments reported by Minkov *et al* [51,52]) was used. By doing this, we used “classical” approach implemented by Uemura [18] to calculate the Fermi temperature, $T_F$, in cuprates from ground state Londoin penetration depth, λ(0). Data points derived by this approach designated in Figures 1-3 by “from $\lambda(0)$”.

2. $T_F$ was calculated based Eqs. 6-9,18, in which utilized $\xi(0)$ values. Data points derived by this approach designated in Figures 1-3 by “from $\xi(0)$”.

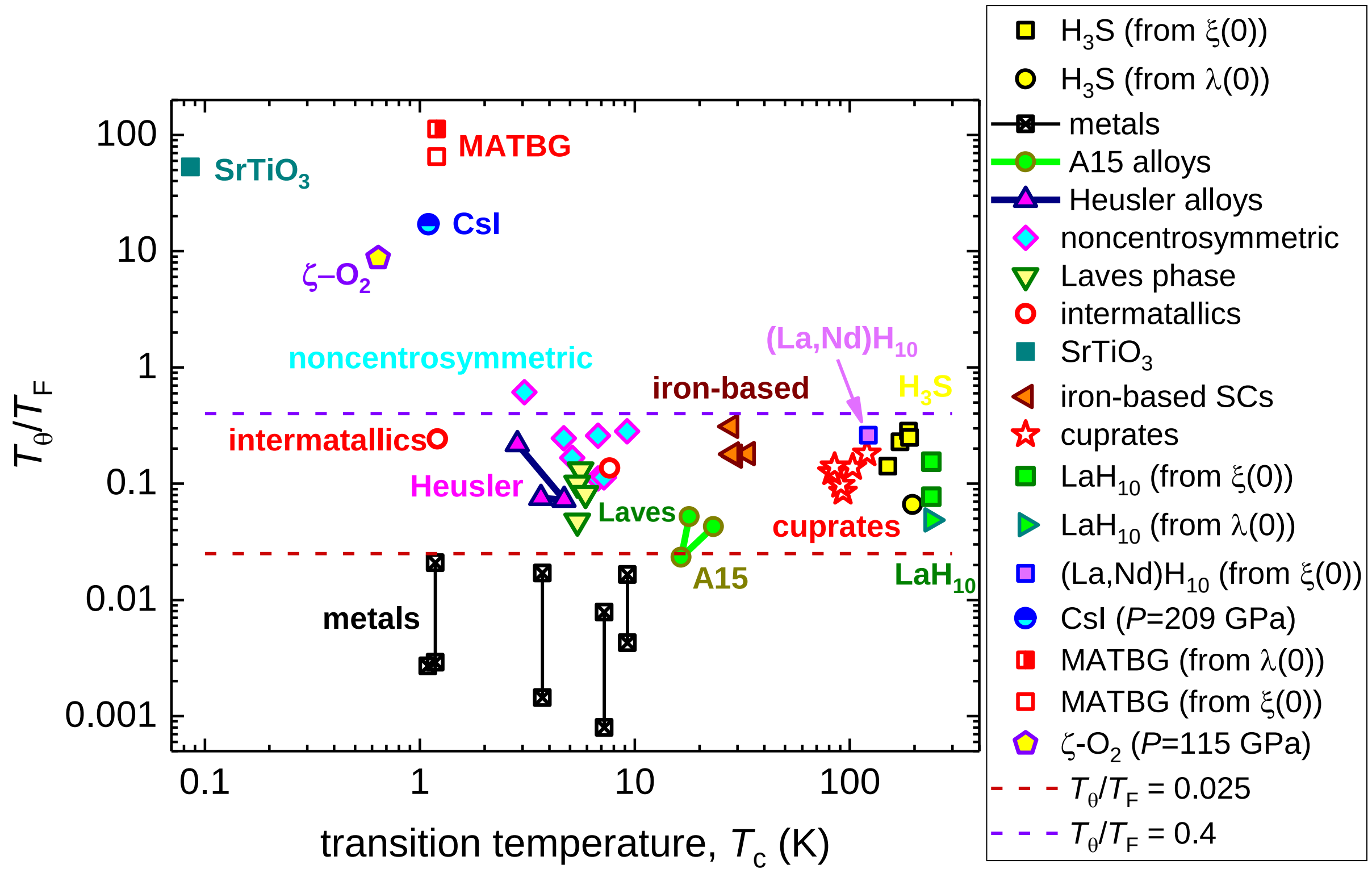

**Figure 3.** Plot $\frac{T_\theta}{T_F}$ vs $T_c$ for primary superconducting families. References on original data ($T_\theta$, $T_c$, $T_F$) can be found in Table 1.

It should be noted that in both approaches the electron-phonon coupling strength constant, $\lambda_{e-ph}$, was assumed to be $\lambda_{e-ph} = 1.76$, which is averaged value of values calculated by first-principles calculations [55,56] and values extracted from experimental $R(T)$ data [27].

It can be seen in Table 1 and Fig. 1, that calculated $T_F$ values for $H_3S$ by two alternative approaches are in a very good agreement with each other. To demonstrate the acceptable level of variation in $T_F$ values for the same material, in Table 1 and Fig. 1 we presented results of calculations for pure metals, where $T_F$ was calculated by two mentioned above approaches and the use of experimental data reported by different research groups.

$T_F$ in HTS cuprates were calculated by the Eqs. 13,15 which do not require the knowledge for the electron-phonon coupling constants, $\lambda_{e-ph}$, despite Ledbetter *et al* [7] reported so-call the effective electron-phonon coupling strength, $\lambda_{e-ph,eff}$, from which the effective mass can be deduced, $m^* = (1 + \lambda_{e-ph,eff}) \times m_e$.

Also, it should be noted that for $YBa_2Cu_3O_7$ Uemura [83] reported the relation [83]:

$$\frac{m^*}{m_e} = 2.5 \quad (17)$$

from which $\lambda_{e-ph} = 1.5$ can be derived. Calculated values are in a reasonable agreement with experimental $\frac{m^*}{m_e}$ values reported by several research groups [84-86] in $YBa_2Cu_3O_{7-x}$.

However, because the phenomenology of the electron-phonon mediated superconductivity cannot describe the superconducting state in cuprates, and the $T_\theta$ for cuprates were taken as experimental values (see, for instance, report by Ledbetter *et al* [7,87]), all cuprate superconductors are shown in Fig. 1,3 and are not shown in Fig. 2.

It should be mention the result of $T_F$ calculation in MATBG (Table 1), $T_F = 16.5\ K$, which was primarily based on the London penetration depth, $\lambda(0) = 1860\ nm$, deduced in Ref. 95 from the self-field critical current density, $J_c(sf,T)$, by the approach proposed by us [88]:

$$J_c(sf,T) = \frac{\phi_0}{4\pi\mu_0}\frac{ln\left(1+\sqrt{2}\kappa(T)\right)}{\lambda^3(T)} \quad (18)$$

Remarkable agreement of the deduced value, $T_F = 16.5\ K$, and the value reported in original work on MATBG by Cao *et al* [94], $T_F = 17\ K$, which was calculated based on normal state charge carriers density in MATBG, independently validates our primary idea [88] about fundamental nature of the self-field critical current in weak-links samples [88,89,104]. Recent prove of this concept has reported by Paturi and Huhtinen [105] who utilized a fact that the London penetration depth, $\lambda(0)$, in real samples depends from the mean free-path of charge carriers, $l$:

$$\lambda(0) = \lambda_{clean\ limit}(0)\sqrt{1+\frac{\xi(0)}{l}} \quad (19)$$

where $\lambda(0)$ is effective penetration depth, and $\lambda_{clean\,limit}(0)$ is the penetration depth in samples exhibits very long mean free-path, $l \gg \xi(0)$. Paturi and Huhtinen [105] varied $l$ in $YBa_2Cu_3O_{7-x}$ films and showed that the change in $J_c(st, T)$ satisfies with Eqs. 18,19.

The MATBG does not show in the Figure 2, because the derivation of $\lambda_{e-ph}$ cannot be performed by the used phenomenology: $m^* = (1 + \lambda_{e-ph}) \times m_e$, because $\frac{m^*}{m_e} = 0.2$ [95], however, this material is shown in Figures 1,3, because $\lambda_{e-ph}$ is not required for these plots.

Returning back to hydrides, we need to note, that Durajski [56] performed first-principles studies on the strength of nonadiabatic effects in highly-compressed sulphur hydride and phosphorus hydride. Calculations [56] showed that the strength of the nonadiabatic effects in these hydrides can be quantify as moderate in comparison with classical nonadiabatic superconductors, like $SrTiO_3$. This is in a good agreement with our result (see Figure 3 and Table 1) that all deduced $\frac{T_\theta}{T_F}$ values for NRTS are within the range of:

$$0.03 \leq \frac{T_\theta}{T_F} \leq 0.3. \qquad (20)$$

Considering primary difference between Figs. 1-3, we need to mention that very strong nonadiabatic superconductors (i.e. MATBG, CsI, and $\zeta$-$O_2$) in the Uemura plot (Fig. 1) are located inside of the unconventional superconductors band $0.01 \leq \frac{T_c}{T_F} \leq 0.05$ which is not a clear way to manifest primary unique property of this materials, because pnictides, cuprates, and hydrides are also located in the same band.

Moreover, classical nonadiabatic superconductor $SrTiO_3$ falls in intermediate zone between unconventional and BCS superconductors, because for this material exhibits $\frac{T_c}{T_F} = 0.0066$, and by this criterion, $SrTiO_3$ is similar to the Laves phase materials, intermetallics, A-15 alloys, and Heusler alloys, which cannot be considered to be correct manifestation of primary uniqueness of this materials which is nonadiabaticity.

However, in Figs. 2 and 3, the outstanding separations of all nonadiabatic superconductors from its adiabatic and moderate nonadiabatic counterparts is clearly manifested.

By looking at the data in Figures 2 and 3, it is easy to recognize that ¾ (30 of 40) of analysed superconductors fall into reasonably narrow band:

$$0.025 \leq \frac{T_\theta}{T_F} \leq 0.4 \qquad (21)$$

despite a fact, that no any selectivity criterion was applied to include data in the databased.

Based on this, we proposed to use the values in the Eq. 21 as empirical limits for the adiabatic superconductors ($\frac{T_\theta}{T_F} \leq 0.025$), moderate nonadiabatic superconductors ($0.025 \leq \frac{T_\theta}{T_F} \leq 0.4$), and strong nonadiabatic superconductors ($\frac{T_\theta}{T_F} \geq 0.4$).

As it follows from our analysis that all strong nonadiabatic superconductors exhibits low superconducting transition temperatures, $T_c \leq 1.2\ K$ (Fig. 3).

It should be mentioned that Hirsch and Marsiglio in series of papers [106-109] claimed the absence of the superconductivity as a physical phenomenon in highly-compressed hydrides. In one occasion [110], we already showed that one their claims [109] is incorrect.

It is important to stress, that in 2015 Hirsch and Marsiglio [111] completely supported the discovery of the superconductivity in $H_3S$ and extended their theory of hole superconductivity [112-114] on $H_3S$ and "other sulphates" [111]. In overall, Hirsch and Marsiglio theory of hole superconductivity [112-114] is (not exactly, but roughly speaking) based on an idea that "only hole conductors can be superconductors" [114].

In following years (after 2015), Hirsch and Marsiglio published several papers [115-117], where in regard of $H_3S$, Souza and Marsiglio stated: "…We find high critical temperatures are possible, even with very modest coupling strengths. …" [116].

More clearly the statement had expressed by Hirsch [117] in 2017:

"*… In addition, we have argued […] that these models lead to hole pairing and superconductivity in the following classes […] of superconducting materials:*

*(1) Hole-doped cuprates […]*

*(2) Electron-doped cuprates […]*

*(3) Magnesium diboride […]*

*(4) Transition metal series alloys […]*

*(5) Iron pnictides […]*

*(6) Iron selenides […]*

*(7) Doped semiconductors […]*

*(8) Elements under high pressure […]*

*(9) Sulphur hydride […]*

*(10) A-15 materials […]*

*(11) All other superconductors […]*".

All claims by Hirsch and Marsiglio [106-109] about the absence of the phenomenon of the superconductivity in hydrides have a very sharp starting time, which is the publication of experimental data by Mozafarri *et al* [54] who showed that the $H_3S$ is the electron-type conductor.

Due to, in overall, Hirsch and Marsiglio theory of hole superconductivity [112-114] is in a strike disagreement with experimental data on cuprates and pnictides [118], and, even, on $SrTiO_3$ (Table 1) (which is *n*-type semiconductor [8]), that eventually all claims by Hirsch and Marsiglio [106-109] about the absence of superconductivity in hydrides will be answered based on scientific background. However, the Brandolini's law [119] is directly applied to the efforts to refute statements expressed by these authors.

## 5. Conclusions

In this work we proposed a new calcification scheme to quantify the effects of nonadiabaticity in superconductors. By performing the analysis of experimental data for 40 superconductors, which represent primary families of superconductors, we found that ¾ of all analysed superconductors falls into narrow $0.025 \leq \frac{T_\theta}{T_F} \leq 0.4$ band. Based on this, we proposed the taxonomy for the strength of the nonadiabatic effects in superconductors.

## Acknowledgments

The author thanks Luciano Pietronero (Universita' di Roma) for explanations of the limitations of the Bardeen-Cooper-Schrieffer (BCS) and Migdal–Eliashberg (ME) theories of electron–phonon-mediated superconductivity. The author also thanks Dmitry V. Semenok (Skolkovo Institute of Science and Technology) for reading and commenting the paper.